\documentclass[%
reprint,
 amsmath,amssymb,
 aps,
]{revtex4-1}

\usepackage{graphicx}
\usepackage{dcolumn}
\usepackage{bm}

\usepackage{color}

\begin{document}


\title{$^{17}$O NMR study of charge ordered La$_{1.885}$Sr$_{0.115}$CuO$_4$}

\author{T.\ Imai} 
\affiliation{Department of Physics and Astronomy, McMaster University, Hamilton, Ontario L8S4M1, Canada}
\affiliation{Canadian Institute for Advanced Research, Toronto, Ontario M5G1Z8, Canada}
\author{K.\ Hirota}\email{Deceased}
\affiliation{Department of Physics, Tohoku University, Aramaki Aoba, Sendai 980-8578, Japan}

\date{\today}

\begin{abstract}
We successfully isolated the $^{17}$O NMR signals at the planar sites in La$_{1.885}$Sr$_{0.115}$CuO$_4$ ($T_{c}=30$~K) by applying an external magnetic field along the a-axis of a single crystal.  We demonstrate that charge order enhances incommensurate Cu spin fluctuations below $T_{charge} \simeq 80$~K.

\end{abstract}

\maketitle

Soon after the discovery of charge \textquotedblleft stripe" order at $T_{charge} = 65$~K in La$_{1.6-x}$Nd$_{0.4}$Sr$_{x}$CuO$_4$ ($x \sim 1/8$) \cite{TranquadaNature1995}, we identified peculiar NMR anomalies at and below the charge order transition \cite{HuntPRL,HuntPRB}: (i) $^{63}$Cu NMR intensity is gradually wiped out below $T_{charge}$, even if we take into account the transverse relaxation effects on the spin echo intensity, (ii) the Gaussian spin echo decay induced by the indirect nuclear spin-spin coupling becomes unobservable, and  (iii) low frequency Cu spin fluctuations as probed by $^{139}$La nuclear spin-lattice relaxation rate $1/T_1$ are enhanced.  We observed all of these peculiar \textquotedblleft NMR fingerprints" of charge order also in  La$_{1.885}$Sr$_{0.115}$CuO$_4$ ($T_{c} =  30$~K), and hence concluded the existence of charge ordered phase in the superconducting cuprate \cite{HuntPRL}.  

Our 1999 conclusion became controversial in the ensuing years, because diffraction experiments  failed to detect the Bragg peaks arising from charge density modulation.   Moreover, a widely shared view at the time was that charge order was a byproduct of the low temperature tetragonal (LTT) structural phase transition at $\sim70$~K in La$_{1.48}$Nd$_{0.4}$Sr$_{0.12}$CuO$_4$.  Superconducting La$_{1.885}$Sr$_{0.115}$CuO$_4$ does not undergo the LTT transition.  Recent advances in the instrumental technologies have finally led to successful detection of charge order Bragg peaks of La$_{1.885}$Sr$_{0.115}$CuO$_4$ at as high as $T_{charge} \simeq 80$~K \cite{Croft, Thampy, He}.  The new revelation also indicated that our original estimation of $T_{charge} \simeq 50$~K \cite{HuntPRL} overlooked the gradual onset of charge order.   

We recently revisited the issue of charge order based on $^{63}$Cu and $^{139}$La NMR using a new single crystal sample of La$_{1.885}$Sr$_{0.115}$CuO$_4$ \cite{ImaiPRB2017, ArsenaultPRB2017}.  Owing to the state-of-the-art NMR technologies, we were able to show that the lost spectral weight of the $^{63}$Cu NMR signals below $T_{charge} \simeq 80$~K reemerges as a very broad, wing-like NMR signal \cite{ImaiPRB2017}.  The relaxation rates of the wing-like signal are so fast that one can detect the NMR signal only with extremely short NMR pulse separation time $\tau \sim 2~\mu$s.  We also demostrated that the volume fraction of the CuO$_2$ planes under the influence of charge order does not reach 100~\% until the temperature is lowered to $\sim 20$~K.  

In this Short Note, we report $^{17}$O single crystal NMR study of La$_{1.885}$Sr$_{0.115}$CuO$_4$.  From early days, accurate $^{17}$O NMR measurements below $\sim 100$~K at the planar $^{17}$O sites in La$_{2-x}$Sr$_{x}$CuO$_4$ have been known to pose a technical challenge, because it is difficult to properly resolve the planar and apical $^{17}$O NMR signals even with a single crystal sample.   Our $^{17}$O single crystal NMR measurements conducted in an external magnetic field $B_{ext}$ applied along the crystal c-axis \cite{SingerPRB2005} covered only the temperature range down to $\sim 100$~K for that very reason.     

\begin{figure}
\includegraphics[width=3in]{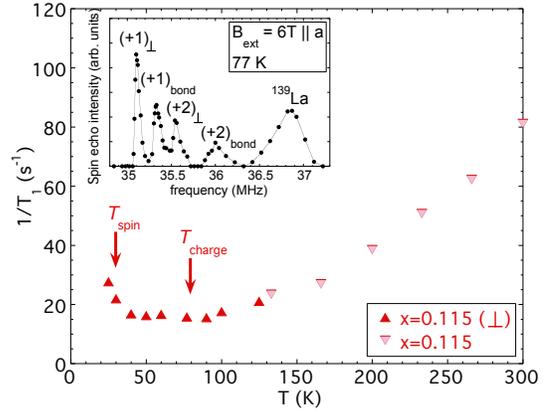}
\caption{(Color online) $1/T_{1}$ of the planar  $^{17}$O sites measured at the (+1)$_{\perp}$ peak ($\blacktriangle$) and the central transition with the $B_{ext}~||~c$ geometry ($\blacktriangledown$).  Inset: $^{17}$O NMR lineshape observed at 77~K in $B_{ext} = 6$~T applied along the a-axis.  The first and second upper satellite peaks are identified with +1 and +2, respectively, whereas the subscript $\perp$ and {\it bond} signify the direction of the magnetic field within the CuO$_2$ plane.  The peak around the 36.8~MHz is the central transition of the $^{139}$La site.  All other $^{17}$O peaks and the lower satellite peaks of $^{139}$La NMR appear below 34.8~MHz.  }
\label{f1}
\end{figure}

Here we show that there is a way to circumvent the superposition problem by using an alternate field configuration, and explore the crucial temperature range below $T_{charge} \simeq 80$~K.  In the inset of Fig.\ 1, we present the $^{17}$O NMR lineshape measured at 77~K in $B_{ext} = 6$~T applied along the a-axis.  In this field geometry, $B_{ext}$ is along the Cu-O-Cu bond for 50~\% of the planar $^{17}$O sites, whereas the remaining 50~\% has the field applied perpendicular ($\perp$) to the Cu-O-Cu bond.  Luckily, the nuclear quadrupole frequency $\nu_{Q}^{(bond)} \simeq 0.66$~MHz and $\nu_{Q}^{(\perp)} \simeq 0.44$~MHz along the bond and perpendicular direction of the planar sites are much larger than  $\nu_{Q}^{(ab)} \simeq 0.10$~MHz for the apical sites, and hence the upper satellite peaks are isolated even below $100$~K.  We found no evidence for quadrupole broadening of $^{17}$O as well as $^{63}$Cu \cite{ImaiPRB2017} and $^{139}$La \cite{ArsenaultPRB2017} NMR peaks below $T_{charge}$, suggesting that the amplitude of charge density modulation is very small.

\begin{figure}
\includegraphics[width=3in]{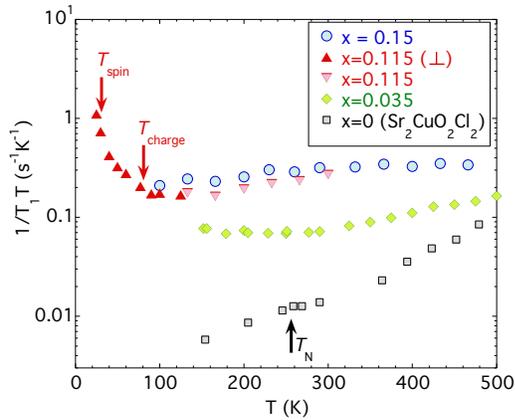}
\caption{(Color online) $1/T_{1}T$ at the planar  $^{17}$O sites in La$_{2-x}$Sr$_{x}$CuO$_4$ with $x=0.115$ (this work), $x=0.15$ \cite{SingerPRB2005}, $x=0.035$ \cite{ThurberPRL1997}, and undoped parent phase Sr$_2$CuO$_2$Cl$_2$ ($T_{N}=257$~K) \cite{ThurberPRL1997}.  All the data except for those marked as \textquotedblleft $\perp$" for $x=0.115$ were measured with the magnetic field applied along the c-axis.}
\label{f2}
\end{figure}

In the main panel of Fig.\ 1, we summarize the temperature dependence of $1/T_1$ for the planar $^{17}$O sites.  $1/T_1$ exhibits a plateau below $\sim 100$~K, followed by strong enhancement below $\sim 35$~K due to the onset of strong short range order.  We note, however, that $1/T_{1}$ does not diverge at the spin ordering temperature $T_{spin} = 30$~K as determined by elastic neutron scattering for this crystal \cite{Kimura} due to the glassy nature of spin ordering \cite{HuntPRB}.  These findings are consistent with our observation for the $^{63}$Cu \cite{ImaiPRB2017} and $^{139}$La \cite{ArsenaultPRB2017} sites.  In Fig.\ 2, we also summarize $1/T_{1}T \propto \Sigma_{{\bf q}} | A({\bf q}) |^{2} Im~ \chi({\bf q}, \omega_{o})$, where $| A({\bf q}) | ^{2} \propto cos^{2} (q_{x}/2)$ is the hyperfine form factor \cite{MMP}, $Im~ \chi({\bf q}, \omega_{o})$ is the imaginary part of the dynamical spin susceptibility, and $\omega_{o}$ is the NMR frequency.  Above $T_{charge}$, $1/T_{1}T$ monotonically decreases with temperature, similarly to the optimally superconducting $x = 0.15$ \cite{SingerPRB2005}.  Moreover, at the phenomenological level \cite{MMP}, $1/T_{1}T$ closely follows the temperature dependence of the uniform spin susceptibility as measured by the NMR Knight shift \cite{SingerPRB2005}, because the antiferromagnetic Cu spin fluctuations are geometrically cancelled out by the hyperfine form factor \cite{MMP}.  It is worth noting that commensurate antiferromagnetic spin fluctuations in undoped CuO$_2$ planes are almost completely filtered out by the form factor, and $1/T_1$ hardly exhibits any noticeable anomaly at the N\'eel transition at $T_{N} = 257$~K, as shown in Fig.\ 2 for Sr$_2$CuO$_2$Cl$_2$ \cite{ThurberPRL1997}.   

Below $T_{charge}$, $1/T_{1}T$ for La$_{1.885}$Sr$_{0.115}$CuO$_4$ begins to grow, leading us to two significant conclusions.  First, charge order triggers the growth of antiferromagnetic spin fluctuations, in agreement with our earlier report that the aforementioned wing-like $^{63}$Cu NMR signals have extremely fast $1/T_{1}$ \cite{ImaiPRB2017}, and $1/T_{1}T$ at the $^{139}$La sites are enhanced below $T_{charge}$ \cite{HuntPRB, ArsenaultPRB2017}.  Second, in view of the fact that the commensurate spin fluctuations are filtered out by the hyperfine form factor, the enhanced spin fluctuations below $T_{charge}$ must be incommensurate.  Our results are consistent with the inelastic neutron scattering measurements with low energy transfer \cite{ETH}.  

{\bf Acknowledgement}: We acquired the NMR data at M.I.T. in 1998 - 1999 with the  support from NSF DMR 96-23858 and 98-08941.   The work at McMaster was supported by NSERC and CIFAR.  
T.\ I.\ expresses his sincere gratitude to K.\ H.\ for his support for our NMR investigation of charge order before his passing in 2010.


\begin{thebibliography}{12}
\bibitem{TranquadaNature1995} J.\ M.\ Tranquada, B.\ J.\ Sternlieb, J.\ D.\ Axe, Y.\ Nakamura, and S.\ Uchida, Nature ${\bf 375}$, 561 (1995).
\bibitem{HuntPRL} A.\ W.\ Hunt, P.\ M.\ Singer, K.\ R.\ Thurber, and T.\ Imai, Phys. Rev. Lett. ${\bf 82}$, 4300 (1999).
\bibitem{HuntPRB} A.\ W.\ Hunt, P.\ M.\ Singer, A.\ F.\ Cederstr\"om, and T.\ Imai, Phys. Rev. B ${\bf 64}$, 134525 (2001).
\bibitem{Croft} T.\ P.\ Croft, C.\ Lester, M.\ S.\ Senn, A.\ Bombardi, and S.\ M.\ Hayden, Phys. Rev. B ${\bf 89}$, 224513 (2014).
\bibitem{Thampy} V.\ Thampy, M.\ P.\ M.\ Dean, N.\ B.\ Christensen, L.\ Steinke, Z.\ Islam, M.\ Oda, M.\ Ido, N.\ Momono, S.\ B.\ Wilkins, and J.\ P.\ Hill, Phys. Rev. B ${\bf 90}$, 100510 (2014).
\bibitem{He} W.\ He, Y.\ S.\ Lee, and M. Fujita, unpublished data (2017).
\bibitem{ImaiPRB2017} T.\ Imai, S.\ K.\ Takahashi, A.\ Arsenault, A.\ Acton, D.\ Lee, W.\ He, Y.\ S.\ Lee, and M.\ Fujita, Phys. Rev. B ${\bf 96}$, 224508 (2017).
\bibitem{ArsenaultPRB2017} A.\ Arsenault, S.\ K.\ Takahashi, T.\ Imai, W.\ He, Y.\ S.\ Lee, and M.\ Fujita, Submitted to Phys. Rev. B.
\bibitem{SingerPRB2005} P.\ M.\ Singer, T.\ Imai, F.\ C.\ Chou, K.\ Hirota, M.\ Takaba,T.\ Kakeshita, H.\ Eisaki, and S.\ Uchida, Phys. Rev. B ${\bf 72}$, 014537 (2005).
\bibitem{ThurberPRL1997} K.\ R.\ Thurber, A.\ W.\ Hunt, T.\ Imai, F.\ C.\ Chou, and Y.\ S.\ Lee, Phys. Rev. Lett. ${\bf 79}$, 171 (1997).
\bibitem{Kimura} H.\ Kimura, K.\ Hirota, H.\ Matsushita, K.\ Yamada, Y.\ Endoh, S.-H.\ Lee, C.\ F.\ Majkrzak, R.\ Erwin, G.\ Shirane, M.\ Greven, Y.\ S.\ Lee, M.\ A.\ Kastner, and R.\ J.\ Birgeneau, Phys. Rev. B ${\bf 59}$, 6517 (1999).
\bibitem{MMP} H.\ Monien, P.\ Monthoux, and D.\ Pines, Phys. Rev. B ${\bf 43}$,275 (1991).
\bibitem{ETH} A.\ T.\ R\o{}mer, J.\ Chang, N.\ B.\ Christensen, B.\ M\ Andersen, K.\ Lefmann, L.\ M\"ahler, J.\ Gavilano, R.\ Gilardi, C.\ Niedermayer, H.\ M.\ R\o{}nnow, A.\ Schneidewind, P.\ Link, M.\ Oda, M.\ Ido, N.\ Momono, and J.\ Mesot, Phys. Rev. B ${\bf 87}$, 144513 (2013).

\end{thebibliography}
\end{document}